\documentclass[twocolumn]{aastex63}
\usepackage{amsmath}

\newcommand{\swift}{\textit{Swift}}
\submitjournal{ApJ}
\received{January 31, 2022}
\accepted{May 27, 2022}
\shorttitle{A Secondary Radio Flare from AT\,2019azh}
\shortauthors{Sfaradi et al.}

\begin{document}

\title{A Late-Time Radio Flare following a Possible Transition in Accretion State in the Tidal Disruption Event AT\,2019azh}

\correspondingauthor{Itai Sfaradi}
\email{itai.sfaradi@mail.huji.ac.il }

\author[0000-0003-0466-3779]{Itai Sfaradi}
\affiliation{Racah Institute of Physics, The Hebrew University of Jerusalem, Jerusalem 91904, Israel}

\author[0000-0002-5936-1156]{Assaf Horesh}
\affiliation{Racah Institute of Physics, The Hebrew University of Jerusalem, Jerusalem 91904, Israel}

\author{Rob Fender}
\affiliation{Astrophysics, Department of Physics, University of Oxford, Keble Road, Oxford OX1 3RH, UK}

\author[0000-0003-3189-9998]{David A. Green}
\affiliation{Astrophysics Group, Cavendish Laboratory, 19 J. J. Thomson Ave., Cambridge CB3 0HE, UK}

\author[0000-0001-7361-0246]{David R. A. Williams}
\affiliation{Jodrell Bank Centre for Astrophysics, School of Physics and Astronomy, The University of Manchester, Manchester, M13 9PL, UK}
\affiliation{Astrophysics, Department of Physics, University of Oxford, Keble Road, Oxford OX1 3RH, UK}

\author[0000-0002-7735-5796]{Joe Bright}
\affiliation{Department of Astronomy, University of California, Berkeley, CA 94720-3411, USA}

\author{Steve Schulze}
\affiliation{The Oskar Klein Centre, Physics Department of Physics, Stockholm University, Albanova University Center, SE 106 91 Stockholm, Sweden}

\begin{abstract}

We report here radio follow-up observations of the optical Tidal Disruption Event (TDE) AT\,2019azh. Previously reported X-ray observations of this TDE showed variability at early times and a dramatic increase in luminosity, by a factor of $\sim 10$, about 8 months after optical discovery. The X-ray emission is mainly dominated by intermediate hard--soft X-rays and is exceptionally soft around the X-ray peak, which is $L_X \sim 10^{43} \rm \, erg \, s^{-1}$. The high cadence $15.5$\,GHz observations reported here show an early rise in radio emission followed by an approximately constant light curve, and a late-time flare. This flare starts roughly at the time of the observed X-ray peak luminosity and reaches its peak about $110$\,days after the peak in the X-ray, and a year after optical discovery. The radio flare peaks at $\nu L_{\nu} \sim 10^{38} \rm \, erg \, s^{-1}$, a factor of two higher than the emission preceding the flare. In light of the late-time radio and X-ray flares, and the X-ray spectral evolution, we speculate a possible transition in the accretion state of this TDE, similar to the observed behavior in black hole X-ray binaries. We compare the radio properties of AT\,2019azh to other known TDEs, and focus on the similarities to the late time radio flare of the TDE ASASSN-15oi.
\end{abstract}

\smallskip
\section{Introduction} 
\label{sec:intro}

A star that passes too close to a super-massive black hole (SMBH) might be torn apart by tidal forces exerted on it by the SMBH \citep{Rees_1988}. A multi-wavelength electromagnetic flare sometimes follows such a Tidal Disruption Event (TDE), offering a unique real-time opportunity to examine processes related to SMBHs and their interaction with their environments (e.g. accretion) even for dormant SMBHs. Despite the growing number of observed TDEs, this phenomenon is still not well understood. Thermal emission detected in optical and UV is associated with the debris of the disrupted star (e.g., \citealt{Rees_1988,Cannizzo_1990,Komossa_2015,Metzger_2016}). However, it is not clear whether this emission is due to accretion onto the SMBH \citep{Rees_1988,Phinney_1989,evans_1989,Mummery_2019a,Mummery_2019b}, or due to some other process, e.g. internal shocks due to collisions in the debris stream \citep{shiokawa_2015,piran_2015,Liptai_2019,Bonnerot_2020}. In rare occasions, relativistic jets are launched, resulting in luminous emission as in the case of SwiftJ1644+57 (\citealt{Bloom_2011,Burrows_2011,Levan_2011,Zauderer_2011}). Broadband observations of TDEs in multiple wavelengths can reveal their nature and answer many of the still open questions.

So far, radio observations of TDEs revealed diverse properties. Broadband radio observations of the TDE SwiftJ1644+57 (\citealt{Zauderer_2011,Berger_2012,Zauderer_2013}) revealed an initially mildly relativistic collimated jet that expanded and slowly decelerated. The temporal evolution of the radio spectra suggested an increase in the total energy by about an order of magnitude, which can be explained by a structured jet \citep{Berger_2012}. An alternative explanation was suggested by \cite{Piran_2013}, where the energy distribution between the magnetic field and the electrons varies with time. Combined late-time radio and X-ray observations of SwiftJ1644+57 point to deviation from equipartition between the fractions of energy deposited in accelerated electrons and magnetic fields, respectively \citep{Eftekhari_2018}.

In addition to relativistic TDEs (e.g. SwiftJ1644+57; \citealt{Zauderer_2011,Berger_2012,Zauderer_2013,cendes_2021_J1644}, SwiftJ2058+05; \citealt{cenko_2013,Pasham_2015,Brown_2017}, SwiftJ112-82; \citealt{Brown_2017}) a different subclass is that of thermal TDEs discovered by their optical/UV emission. ASASSN-14li \citep{Holoien_2016a,Bright_2018}, one of the well-studied thermal TDEs, showed prompt radio emission, orders of magnitude fainter than that observed in SwiftJ1644+57. \cite{Alexander_2016} interpreted this radio emission as arising from a sub-relativistic accretion-driven wind interacting with the circum-nuclear material (CNM), while \cite{Krolik_2016} suggested that the outflow is unbound stellar debris traveling away from the SMBH. \cite{van_velzen_2016} on the other hand, attributed this radio emission to a newly launched narrow jet.

Radio discoveries of TDEs are increasing in numbers in recent years, revealing exciting new features. AT\,2019dsg, a TDE which exhibited radio emission similar to ASASSN-14li, was associated with a neutrino emission \citep{Stein_2021}, although this association is still debated in the community (see e.g. \citealt{cendes_2021_dsg}). Very long baseline radio observations of Arp 299B-AT1 \citep{Mattila_2018}, a TDE candidate discovered as an infra-red flare, revealed a long-lived radio jet. While detections of TDEs in radio wavelengths reveal a wide range of properties, a large fraction of TDEs were not detected at all, suggesting a more complicated physical picture \citep{Alexander_2020}. A new development is the discovery of a delayed (by months and years after optical discovery) radio flare from the TDE ASASSN-15oi \citep{Horesh_2021a}. According to \cite{Horesh_2021a} the delayed radio emission cannot be explained by standard models, such as the ones used to describe the radio emission from either SwiftJ1644+57 or ASASSN-14li. The delay may indicate a late-time outflow, possibly due to a transition in the SMBH accretion state. Recently two other examples of possible delayed radio flares in TDEs were reported: iPTF\,16fnl \citep{Horesh_2021b} and IGRJ\,12580+0134 \citep{perlman_2021}, suggesting this may be a common, but mostly unexplored, late-time phase in TDEs. 

Accreting black hole systems such as X-ray binaries (XRBs) and Active Galactic Nuclei (AGNs) also exhibit radio flares. In XRBs, a radio flare, following an X-ray flare and a spectral transition (from a hard to a soft X-ray state), is usually associated with a transition in accretion states \citep{Fender_2004}. A similar transition in accretion states has been suggested to occur in AGNs, based on a statistical analysis of large samples (hundreds to thousands of AGNs; \citealt{kording_2006, Svoboda_2017, fernandez_2021}). Recent observations also find an increasing number of AGNs that are transitioning from being radio-quiet to radio-loud (as their radio luminosity increases dramatically over long time scales). These changes in radio brightness are believed to be the signature of newly launched jets in those AGNs (although it is still unclear whether this is related to accretion state transition; \citealt{Kunert_2020,Nyland_2020}). In TDEs, a transition in the accretion state has been suggested to separately explain the spectral evolution in the X-ray emission of AT\,2018fyk \citep{wevers_2021}, and the delayed radio flare in ASASSN-15oi \citep{Horesh_2021a}. Some theories suggest that this type of transition will be accompanied by a formation of relativistic jets \citep{giannios_2011}. However, a combined accretion state transition signature of a radio flare following an X-ray flare has never been observed in TDEs until now. Here, we report for the first time such a signature, observed in the TDE AT\,2019azh. 

In the following work, we present new radio observations of the TDE AT\,2019azh and analyze them in combination with the X-ray data reported by \cite{Hinkle_2021}. In \S\ref{sec: observations} we present the radio observations we obtained for this TDE. We analyze the temporal evolution of the radio light curve of AT\,2019azh, and the observed late time radio flare (see \S\ref{sec: modeling_and_analysis}). In \S\ref{sec: discussion} we discuss the radio and X-ray connection and the possible transition in accretion state in light of the late time radio flare (\S\ref{subsec: xray}), and compare the radio light curve of AT\,2019azh to light curves of other well-studied TDEs (\S\ref{subsec: compare_TDEs}). \S\ref{sec: conclusions} is for conclusions.

\section{Observations}
\label{sec: observations}

\subsection{Optical Discovery by ASAS-SN and ZTF}
\label{subsec: optical_obs}

The TDE AT\,2019azh (ASASSN-19dj; ZTF17aaazdba) is located at the center of the galaxy KUG 0180+227 at $z=0.022$, with a luminosity distance of $96$\,Mpc \citep{Adelman-McCarthy_2006}. The All Sky Automated Survey for SuperNovae (ASAS-SN) first reported the detection of AT\,2019azh with a magnitude of $16.3$ in $g$-band on February 22.03, 2019 \citep{Brimacombe_ATel,Stanek_TNS}. However, it was also discovered, with a magnitude of $18.99$ in $g$-band, by the Zwicky Transient Facility (ZTF; \citealt{bellm,graham,Dekany_2020}) on February 12.40, 2019 \citep{van_velzen_ATel}. Thus, we set the discovery date to be on MJD $58526.4$. AT\,2019azh was classified as a TDE by \cite{van_velzen_ATel} based on the persistent blue color and spectrum which lacks the features associated with a supernova or an AGN, the high black-body temperature, and the location in the center of its host.

\subsection{Radio observations}
\label{subsec: radio_obs}

Radio observations of AT\,2019azh have been reported by \cite{Perez_ATel}. They reported two separate detections with the electronic Multi-Element Remotely Linked Interferometer Network (eMERLIN\footnote{http://www.e-merlin.ac.uk/}), using a $512$\,MHz band around the central frequency $5.075$\,GHz. These two observations, taken on May 21 and June 11, 2019, resulted in the detection of a radio source with a flux density of $0.35$ and $0.58$ mJy, respectively. These flux densities correspond to luminosity densities of $3.9$ and $6.4 \times 10^{27} \, \rm erg \, s^{-1} \, Hz^{-1}$, respectively. A newly published work by \cite{goodwin_2022} presents a set of multi-wavelength radio observations of this TDE, obtained by the Very Large Array (VLA), spanning from a month up to $\sim 2.5$ years after the optical discovery (however, their data is much more sparse compared to our AMI data; see below).

We conducted an extensive monitoring campaign of AT\,2019azh at a central frequency of $15.5$\,GHz using the Arcminute Microkelvin Imager -- Large Array (AMI-LA; \citealt{zwart_2008}; \citealt{hickish_2018}). Our observing campaign began on February 25, 2019, about $14$\,days after the optical discovery. We observed the source at an high cadence (an observation every week on average), with the last observation undertaken on March 17, 2020. While our first observation resulted in a non-detection (with a $3\sigma$ limit of $F_{\nu} < 0.16$\,mJy; where $F_{\nu}$ is the flux density), we detected a radio source at the position of AT\,2019azh in all of our following observations, with the first detection on March 5, 2019. The flux density at each observation is reported in Table~\ref{tab:AMI_Observations}.

AMI-LA is a radio interferometer comprised of eight, 12.8-m diameter, antennas producing 28 baselines that extend from 18-m up to 110-m in length and operate with a 5 GHz bandwidth, divided into eight channels, around a central frequency of 15.5 GHz. This results in a relatively large synthesized beam of $\sim 30$\,arcsec. Initial data reduction, flagging and calibration of the phase and flux, were carried out using $\tt{reduce \_ dc}$, a customized AMI-LA data reduction software package (e.g.\ \citealt{perrott_2013}). Phase calibration was conducted using short interleaved observations of J0823+2223, while daily observations of 3C286 were used for absolute flux calibration. Additional flagging was performed using CASA \citep{mcmullin_2007}. Images of the field of AT\,2019azh were produced using CASA task CLEAN in an interactive mode. We fitted the source in the phase center of the images with the CASA task IMFIT, and calculated the image rms with the CASA task IMSTAT. We estimate the error of the peak flux density to be a quadratic sum of the image rms, the error produced by CASA task IMFIT, and $5$\,\% calibration error.

\startlongtable
\begin{deluxetable}{cccc}
\tablecaption{AT\,2019azh -- radio observations. \label{tab:AMI_Observations}}
\tablehead{
\colhead{$\Delta t$} & \colhead{$F_{\nu}$} & \colhead{$\Delta F_{\nu}$} & \colhead{Image RMS} \\
\colhead{$[\textrm{Days}]$} & \colhead{$[\textrm{mJy}]$} & \colhead{$[\textrm{mJy}]$} & \colhead{$[\textrm{mJy}]$}
}
\startdata
$14$ & $< 0.16$ & $0.16$ & $0.052$ \\ [0.1ex]
$21$ & $0.22$ & $0.05$ & $0.05$ \\ [0.1ex]
$36$ & $0.18$ & $0.03$ & $0.03$ \\ [0.1ex]
\enddata
\tablecomments{\footnotesize A summary of the $15.5$\,GHz observations of AT\,2019azh conducted with the AMI-LA. $\Delta t$ is the time since optical discovery. See the online table for the full data set.}
\end{deluxetable}

\section{A Late-Time Radio Flare}
\label{sec: modeling_and_analysis}

\begin{figure}
\begin{center}
    \includegraphics[width=0.4\textwidth]{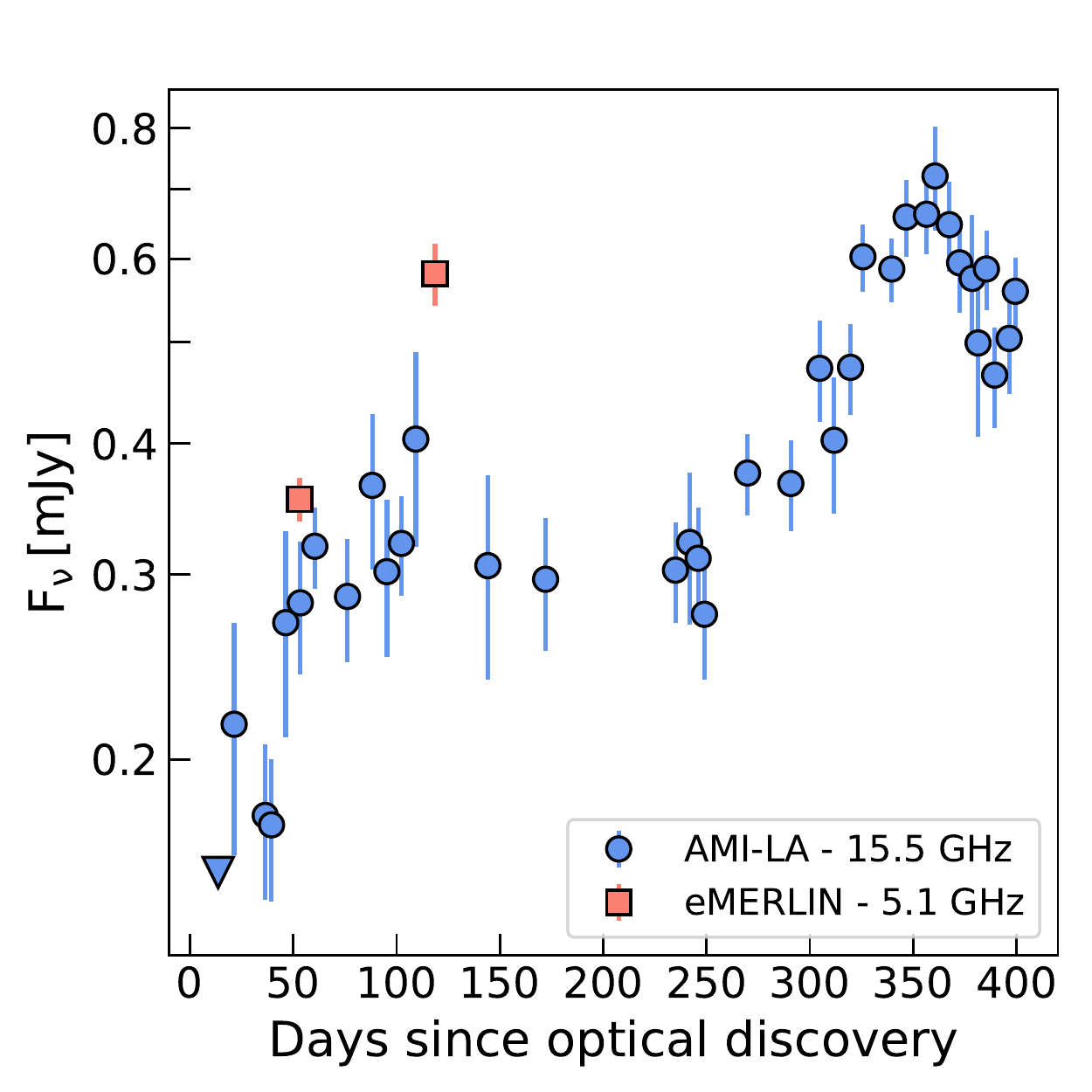}
\end{center}
\caption{\footnotesize{Radio emission from AT\,2019azh as observed by AMI-LA ($15.5$\,GHz; blue circles) and eMERLIN ($5.1$\,GHz; red squares), and reported in \S\ref{subsec: radio_obs}.\label{fig:AT2019azh_radio}}}
\end{figure}

Our $15.5$ GHz observations of AT\,2019azh revealed a peculiar evolution of the radio light curve (see Figure \ref{fig:AT2019azh_radio}). First, the radio emission increases for $100$ days and stays on a plateau ($\sim 0.3$ mJy) for another $170$ days. At $270$ days after the discovery of the TDE, the flux density continues to increase reaching a maximum of $\sim 0.7$ mJy at $360$ days after the discovery. After that the light curve faded. The detailed radio spectrum seen in \cite{goodwin_2022} before and during this radio flare is optically thin at our observed frequency\footnote{We do not make use of the radio data published by \cite{goodwin_2022} since we focus on the temporal evolution of the radio flare which they do not capture due to the lack of high cadence observations.}. Furthermore, the e-MERLIN data suggests that the early emission is already optically thin (see Figure~\ref{fig:AT2019azh_radio}). Therefore, we assume that the emission stays optically thin throughout the late-time flare. If this assumption is true, then the peak of the flare is not due to a spectral turnover (as in, for example, ASASSN-14li; \citealt{Alexander_2016, Krolik_2016}). The latter and the fact that underlying diffuse emission from the host might exist, contaminating the observed radio emission (up to a level of our first non-detection, $\sim 0.16$\,mJy), make it difficult to derive the physical properties of the observed flare (such as the radius of the emitting region, a property which is derived using information on the turnover from an optically thick to an optically thin emission; e.g., \citealt{chevalier_1998}). Nonetheless, we first characterize the temporal behavior of the radio emission and then analyze it with respect to the observed X-ray emission. 

Modeling the radio light curve can be done by introducing a parameterized model, analogous to the one shown in Eq. 4 in \cite{chevalier_1998}. In such a model the flux density, $F_{\nu} \left( \Delta t \right)$, at any time $\Delta t$ after the initial time is given by 
\begin{align}
    \label{eq: parameterized model}
    F_{\nu} \left( \Delta t \right) = & 1.582 \, F_{\nu} \left( t_c \right) \left( \frac{\Delta t}{t_c} \right)^a \times \\
    \nonumber & \left( 1 - e^{- \left( \frac{\Delta t}{t_c} \right)^{-\left( a+b \right)}} \right),
\end{align}
where $t_c$ is the time of the turnover and $F_{\nu} \left( t_c \right)$ is the flux density at this time. The power law of the rising regime of the light curve is given by $a$, and the power law on the declining regime is described by $b$.

\begin{figure}
\begin{center}
    \includegraphics[width=0.4\textwidth]{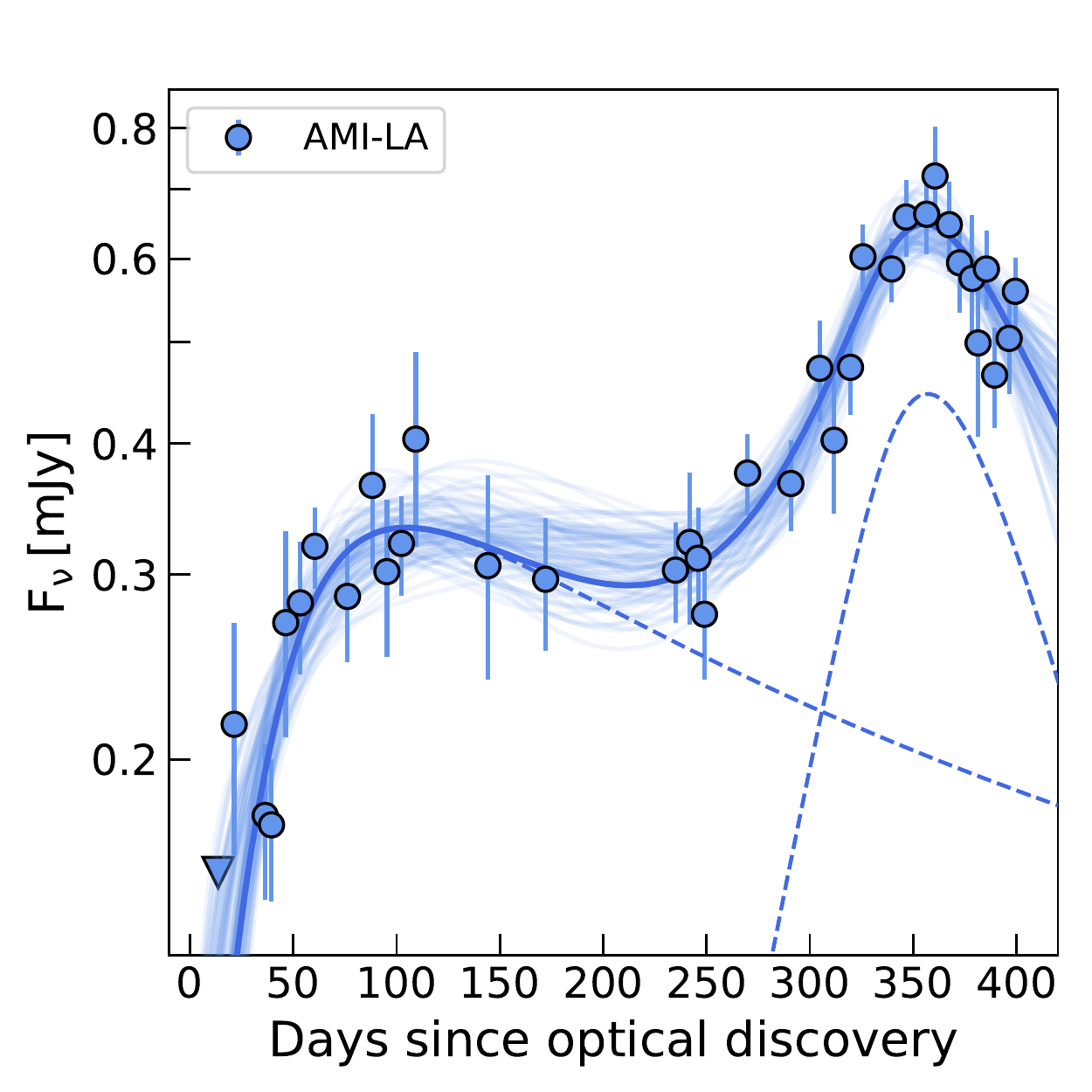}
\end{center}
\caption{\footnotesize{Temporal modeling of the radio emission as two components of synchrotron emitting sources. The dashed blue lines are the two different components, and the solid blue line is the combination of the two. $1\sigma$ uncertainties of the fitting process, drawn from the posterior distributions, are also plotted. \label{fig: temporal_fit}}}
\end{figure}

The early radio emission showing a moderate rise before plateauing, can be interpreted as an initial radio flare due to, for example, an outflow interacting with the CNM (as in ASASSN-14li), or some other emission mechanism, which may be disconnected from the origin of the late-time flare. Thus, we can employ a more comprehensive model where we model the full radio light curve as a combination of two components, where each component (denoted as $1$ and $2$) is described by Eq.~\ref{eq: parameterized model} but with its own set of parameters. We thus fit such a model to the $15.5$\,GHz light curve. The free parameters are the peak flux density, its time, and the temporal power-laws of the two regimes, for each component. We use \texttt{emcee} \citep{foreman_2013} to perform an MCMC analysis and determine the posteriors of the fitted parameters (using flat priors). Our MCMC fitting analysis results in $F_{\nu,1} \left( t_{c,1} \right) = 0.31 \, ^{+0.02} _{-0.03}$\,mJy at $t_{c,1} = 130 \, ^{+60} _{-50}$ days after optical discovery, $a_1 = 0.7 ^{+0.4} _{-0.2}$, and $b_1 = 0.7 ^{+0.5} _{-0.3}$. The second late-time peak is $F_{\nu,2} \left( t_{c,2} \right) = 0.41 \pm 0.06$\,mJy at $t_{c,2}=360 \pm 20$ and we also find $a_2 = 8 ^{+4} _{-3}$ and $b_2 = 6 \pm 2$ (see Figure \ref{fig: temporal_fit}).

\section{Discussion}
\label{sec: discussion}

\subsection{A Combined Radio and X-ray View}
\label{subsec: xray}

\begin{figure*}
\plotone{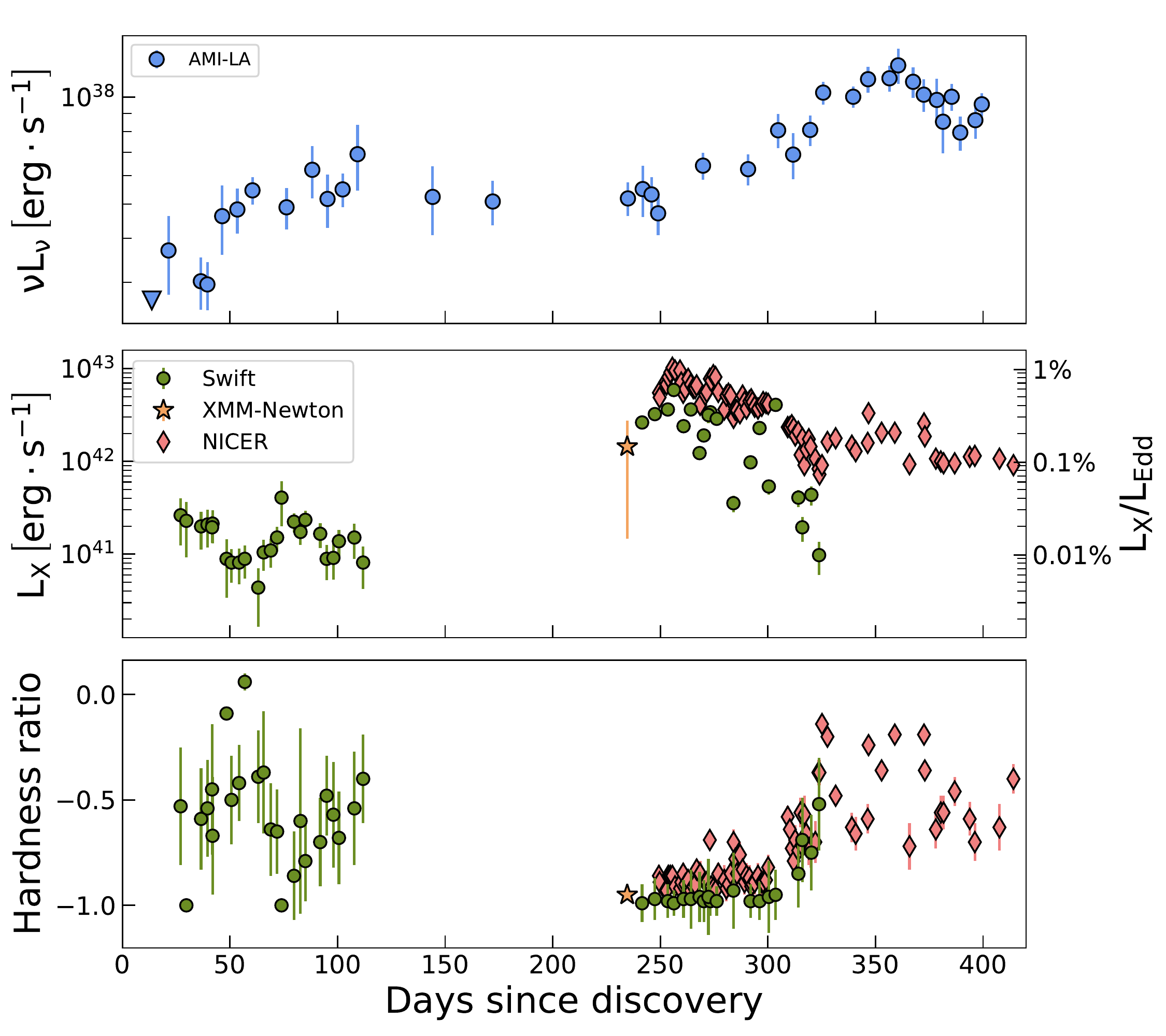}
\caption{\footnotesize{A comparison between the radio light curve (AMI-LA 15.5\,GHz; top panel), the X-ray light curve (\swift/XRT, XMM-Newton, and NICER telescopes; middle panel), and the X-ray hardness ratio temporal evolution (bottom panel). The X-ray plots are reconstructed based on data from \cite{Hinkle_2021} and are similar to some of the panels their Figure 13. The hardness ratio is defined as $\rm HR = \frac{H-S}{H+S}$, where S is the counts in the soft part of the X-ray spectrum ($0.3-2$\,keV), and H is the counts in the hard part of the X-ray spectrum ($2-10$\,keV). \label{fig:X_ray}}}
\end{figure*}

We now discuss the properties of the X-ray light curve obtained with \swift/XRT, XMM-Newton, and NICER ($0.3-10$ keV) as reported in \cite{Hinkle_2021}, and compare it with the $15.5$\,GHz light curve obtained by AMI-LA. The X-ray light curve (Figure \ref{fig:X_ray}) shows early time variations on scales of $10^{41} \rm \, erg \, s^{-1}$ up to $\Delta t \sim 110$\,days ($\Delta t$ is the time since optical discovery), followed by a later flare. The X-ray flare peaks at $\Delta t = 256$\,days with a maximum luminosity of $\sim 10^{43} \rm \, erg \, s^{-1}$, higher by more than an order of magnitude compared to the early emission observed up to $\Delta t = 110$\,days. Since the source became Sun-constrained, there is a seasonal gap in the X-ray data between $\Delta t = 111$ and $234$\,days, thus the evolution of the X-ray light curve leading to the late-time rebrightening is not constrained. Adopting a central SMBH mass of $\rm{M_{BH}} = 7.8 \times 10^6 \, \rm{M_{\odot}}$ \citep{Hinkle_2021}, the $0.3-10$ keV emission in the first $\sim 100$ days is $\sim 0.01\%$ of the Eddington luminosity. The late-time $0.3-10$ keV flare emission peaks at $\sim 1\%$ of the Eddington luminosity\footnote{The Eddington ratio $L_{bol}/L_{EDD}$ is defined as the ratio between the bolometric luminosty and the Eddington luminosity. We approximate the bolometric luminosity by the X-ray luminosity in the $0.3-10$ keV band, as provided by \cite{Hinkle_2021}.} and then declines and settles at a level of $\sim 0.1\%$ Eddington luminosity. The $15.5$\,GHz flare observed with AMI-LA (see Figure \ref{fig:AT2019azh_radio}) immediately follows the X-ray peak emission, and peaks at about $\sim 100$\,days later, when the X-ray emission decreased by an order of magnitude.

\begin{figure*}
\begin{center}
    \includegraphics[width=0.48\textwidth]{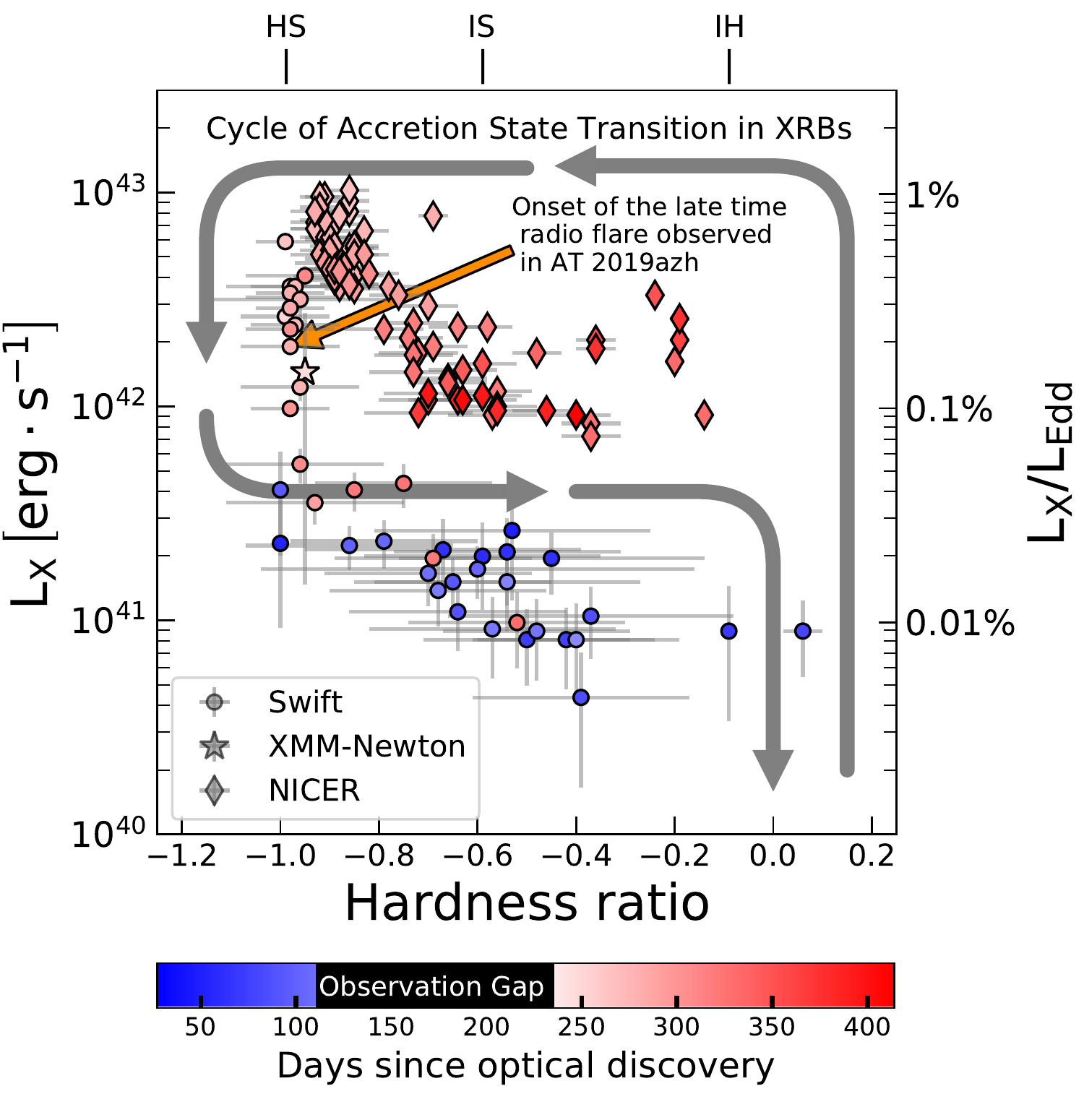}
    \includegraphics[width=0.48\textwidth]{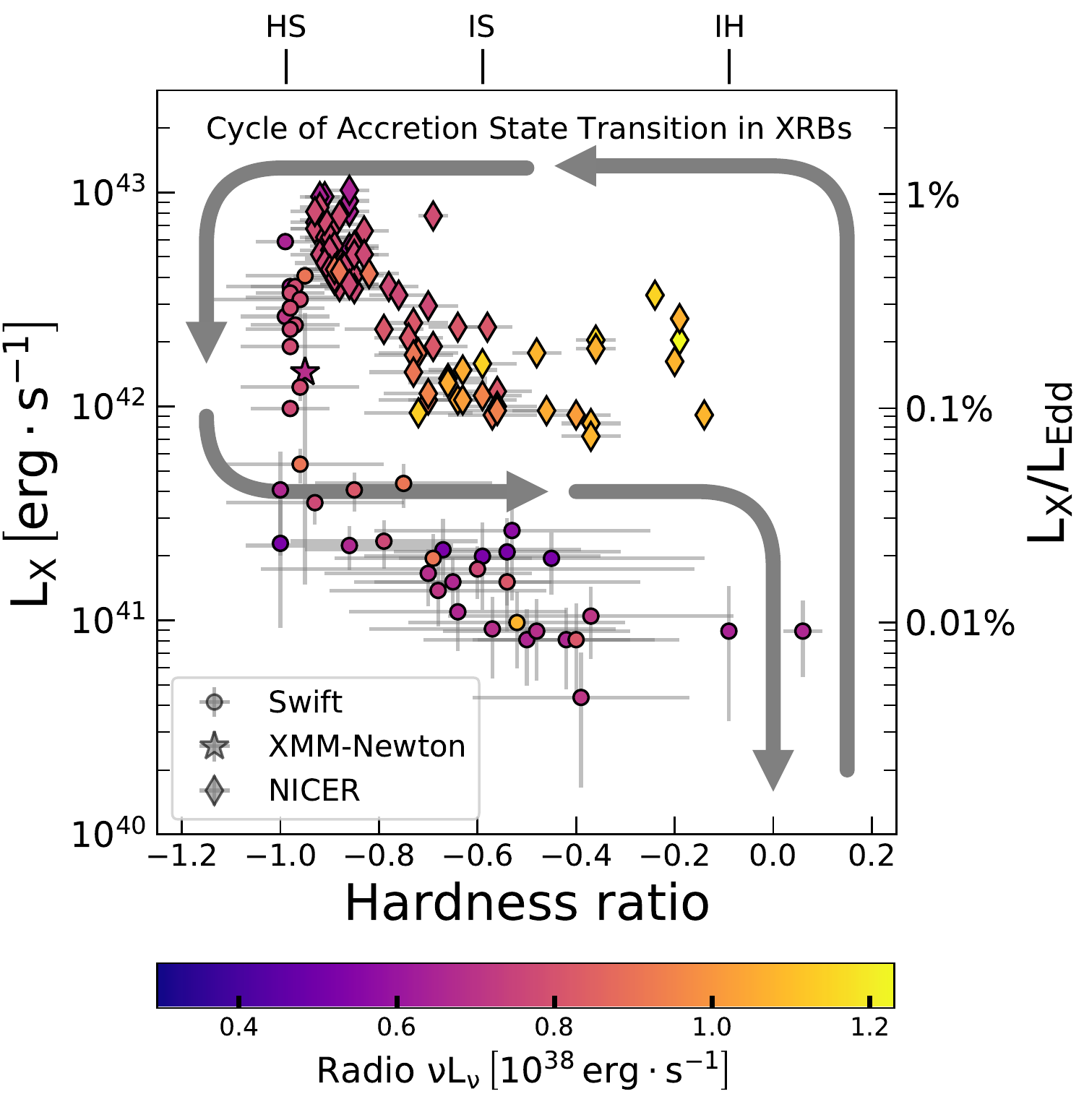}
\end{center}
\caption{\footnotesize{AT\,2019azh X-ray luminosity as a function of its hardness ratio as observed by \swift/XRT, NICER, and XMM-Newton and reported by \cite{Hinkle_2021}. The data on the left panel are color-coded by the observation time, and on the right panel, they are color-coded by the closest observed radio luminosity in the $15.5$ GHz band. Arrows showing the general schematic behavior of the disc-jet coupling observed in XRBs \citep{Fender_2004} are plotted for reference. Also marked on the left panel is the start time of the observed $15.5$ GHz flare. Due to a large gap in the X-ray observation, we do not observe the rise of the X-ray luminosity in the intermediate state, and we observe only the high-soft state (HS) and the transition back through intermediate-soft (IS) towards the intermediate-hard (IH) state. As seen from the right panel, the radio flare observed with AMI-LA follows the soft state and accompanies the transition towards the IH state. }}
\label{fig:transition}
\end{figure*}

The X-ray flare is also accompanied by a significant change in the X-ray emission hardness ratio. The latter is defined as $\rm HR = \frac{H-S}{H+S}$, where H is the counts in the hard part of the spectrum ($2-10$\,keV), and S is the counts in the soft part of the spectrum ($0.3-2$\,keV). This ratio is a measure of the dominant component in the X-ray spectrum, where $\rm HR = 1$ represents a spectrum dominated entirely by hard X-rays, and $\rm HR = -1$ represents a spectrum dominated entirely by soft (thermal) X-rays\footnote{We note that changes in the hardness ratio can also be a result of variations in the absorption column density (due to inflows/outflows around the black hole, for instance) and not necessarily intrinsic.} AT\,2019azh initially features an X-ray spectrum with an intermediate hardness ratio (variable around ${\rm HR}\sim -0.5$) which then completely softens (${\rm HR} \approx -1$) once the X-ray emission flare-up and increases in brightness by more than an order of magnitude. The emission stays soft for more than $50$\,days following the X-ray peak emission and then returns to the intermediate hard-soft state as the X-ray luminosity drops (see bottom panel of Figure \ref{fig:X_ray}).

The behavior of this TDE at late times, in X-ray\footnote{\cite{liu_19azh} suggest that the late-time X-ray behaviour in AT\,2019azh is due to late-time change in accretion.} and radio combined, is reminiscent of the observed behavior of an accretion state transition in XRBs (\citealt{Fender_2004}). This transition occurs when the accretion disc moves in closer to the black hole. This produces an X-ray flare which is followed by a transition from a hard to a soft state. A radio flare, corresponding to a discrete relativistic ejection, is associated with this hard-to-soft state transition. The full cycle of a state transition in an XRB is illustrated in Figure.~\ref{fig:transition}. It appears, however, that the partially hard to soft transition observed in AT\,2019azh does not fully follow the state transition observed in XRBs. This difference may be real or due to the observational gap in the data. 

While in the high-soft (HS) state (and before the cycle continues to an intermediate-soft (IS) state and then to an intermediate-hard (IH) quiescent state) XRBs may exhibit several rapid changes (a mini-cycle) between soft and intermediate hard states, while the X-ray emission remains at relatively high brightness. During these rapid changes, radio flares may also occur \citep{Fender_2004}. Examining the late-time X-ray data, following the soft state, suggests that we are observing a phase similar to the XRB rapid transition change phase, as the X-ray emission spectral state is highly variable and its brightness somewhat decreased but is still significantly brighter than the X-ray emission level shortly after the TDE was discovered. While the time-scale of the X-ray spectral cycle and re-brightening is on the order of at least $50$ days for this TDE, for XRBs the cycles are at an order of hours to approximately a day at maximum. Furthermore, the X-ray luminosity observed in this TDE is significantly higher (few orders of magnitude) than typically observed for XRBs. The several orders of magnitude difference in time-scales and luminosities can be a result of the difference in the system size and the black hole mass, however, the black hole masses of different XRBs are similar and scaling these properties with mass is not trivial.

The behavior of the radio emission with respect to the X-ray emission is somewhat puzzling. First, it is unclear whether the initial radio emission observed shortly after optical discovery is related to an accretion (or transition of its state) onto the SMBH. For instance, it can be a different synchrotron component originating from the interaction of an outgoing stellar debris with the CNM (similar to ASASSN-14li; \citealt{Krolik_2016}). The flaring of the radio emission following the rebrightening of the X-ray emission is reminiscent of the transient radio emission observed in XRBs. However, in XRBs, this transient emission is the result of a relativistic ejection into an existing jet. Following this ejection the production of the jet ceases, the radio emission diminishes, and only then the X-ray becomes dominates by soft (disk) emission. In AT\,2019azh, the late-time radio flare occurs when the X-ray emission is already completely soft. While this is at odds with the transient emission observed in XRBs, the late-time flare we observed might represent a different stage in the XRB cycle, the reemergence of a steady jet. This could have been tested with detailed X-ray observations while the TDE was Sun-constrained and on later times (combined with high cadence radio observations).

It has also been suggested that accretion state transition, such as observed in XRBs, occurs in a similar manner in AGNs \citep{Marscher_2002}. There are several studies \citep{kording_2006, Svoboda_2017, fernandez_2021} analyzing statistical samples of AGNs and matching the various AGN to the different stages of the transition. Eventually, they characterized the state transition by defining three stages in their luminosity vs spectral state phase space. Examining Figure 11 in \cite{kording_2006} suggests that the combined radio and X-ray late-time flaring observed in AT\,2019azh may be comparable with an AGN transitioning from ``stage 3''  (populated by radio-quiet quasars in a soft state) to ``stage 1'' (populated by radio-loud quasars in an intermediate hard state). 

A comparison between radio flares observed in different classes of transients was published by \cite{pietka_2014}. In their Figure 3 they present the radio rise time and luminosity of the different classes. The observed rise time and luminosity of the radio light curve of AT\,2019azh matches to the observed rise time and luminosity of some of the low luminosity AGNs. Note also that the timescale of the radio flare in AT\,2019azh is similar to the one in the AGN observed by \cite{King_2016}, for which high resolution VLBA observations resolved a discrete knot ejection (we note that future high resolution radio observations may reveal such a discrete knot ejection in future nearby TDEs). XRBs, on the other hand, are far less luminous (by $\sim 8$ orders of magnitude) than the observed peak radio luminosity for this TDE, and their typical rise of the radio flare (which is on a scale of a day), is much shorter than the rise time observed for AT\,2019azh ($\sim 100$ days).

\subsection{A Comparison With Other TDEs}
\label{subsec: compare_TDEs}

\begin{figure}
\plotone{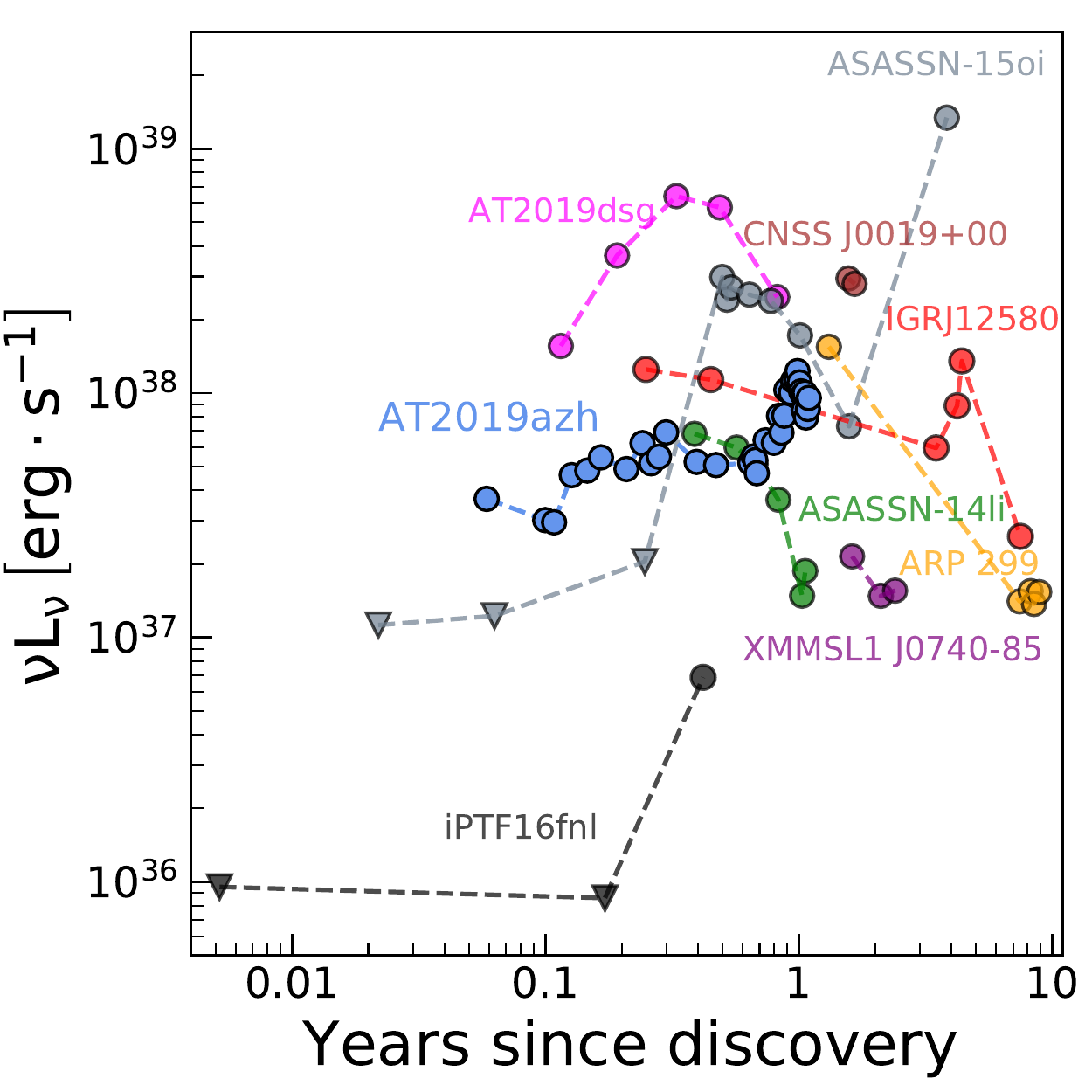}
\caption{\footnotesize{Comparison of the radio light curve of AT\,2019azh to other well-studied TDEs, ASASSN-14li ($5$\,GHz; \citealt{Alexander_2016}), AT2019dsg ($8.49$\,GHz; \citealt{Stein_2021}, $9$\,GHz; \citealt{cendes_2021_dsg}), XMMSL1 J0740-85 ($5.5$\,GHz; \citealt{Alexander_2017}), ARP 299 ($8.4$\,GHz; \citealt{Mattila_2018}), CNSS J0019+00 ($5.1$\,GHz; \citealt{Anderson_2020}), ASASSN-15oi ($5$\,GHz; \citealt{Horesh_2021a}), IGRJ12580+0134 ($1.5$\,GHz; \citealt{perlman_2021}), and iPTF\,16fnl ($6$\,GHz; \citealt{Horesh_2021b}). Due to its high luminosity (two orders of magnitude higher than this plot), and relativistic nature of the event, we do not present Swift J1644 in this plot.\label{fig:comparison}}}
\end{figure}

Past radio observations of TDEs revealed diverse properties (Figure \ref{fig:comparison} presents a comparison between our radio measurements of AT\,2019azh, and past radio-detected TDEs). Until recently, in most of the radio detected events, the radio emission has been explained as originating from the interaction of an outflow (of different origins), launched promptly after stellar disruption, with the CNM. Relativistic TDEs, such as SwiftJ1644 \citep{Zauderer_2011,Berger_2012,Eftekhari_2018} usually produce peak luminosities orders of magnitudes higher than the radio peak observed in the late-time flare of AT\,2019azh. On the other hand, the early radio emission of AT\,2019azh is at a similar flux density level to that of ASASSN-14li\footnote{While ASASSN-14li did not exhibit a delayed radio (or X-ray) flare, \cite{Pasham_2018} claim that there is a lagged correlation, of about $12$ days, between low-level variability in the X-ray emission and the radio emission.}. In the latter, the radio emission is suggested to originate from a subrelativistic outflow (the nature of which is still debated) with the CNM. Moreover, in contrast to AT\,2019azh, late time observations of ASASSN-14li \citep{Bright_2018} showed no late time radio rebrightening (note however that their data has a gap between 170 days to 390 days after optical discovery).

Late-time (delayed) radio flares have been observed so far only in three cases (ASASSN-15oi; \citealt{Horesh_2021a}, iPTF\,16fnl; \citealt{Horesh_2021b}, and IGR\,J12580+0134; \citealt{perlman_2021}). It is worth noting that the timescale of the onset of the late-time radio flare in the first two events is similar to the one of AT\,2019azh at around $\sim 200$\,days (surprisingly enough, this is roughly the timescale on which a neutrino was observed in association with the TDE AT\,2019dsg; \citealt{Stein_2021}). In terms of luminosity, AT\,2019azh is more luminous than iPTF\,16fnl by an order of magnitude and less luminous than ASASSN-15oi by a factor of $\sim 3$. The similarity between the events may also extend into the X-ray regime. The X-ray emission from ASASSN-15oi increased beyond $1\%$ Eddington luminosity at late times, and its thermal component became brighter \citep{Horesh_2021a}. However, a large gap in the observed X-ray data during the radio flare made any combined X-ray--radio analysis limited. Late-time X-ray rebrightening was also observed in the TDE candidate AT\,2018fyk \citep{wevers_2021}, but it was not accompanied by a radio flare. 

\section{Conclusions and Summary}
\label{sec: conclusions}

In this work, we report a comprehensive set of $15.5$\,GHz AMI-LA observations of AT\,2019azh. These observations span more than a year after optical discovery and show variability on different time scales. The first few measurements exhibit a rise of the flux density until reaching an approximately  constant level of emission of $\sim 0.3$\,mJy. At $\sim 270$\,days after the optical discovery, a late-time radio flare is evident when the flux density rises again and reaches a peak flux density of about a factor of two higher at $\sim 360$\,days. A multi-wavelength campaign in the radio, conducted with the VLA \citep{goodwin_2022}, does not capture this late time flaring of the radio emission due to low-cadence observations. However, their radio data suggests additional flaring of the radio emission at late-times (after the flare observed by us).

A similar behavior of the X-ray emission is observed and reported by \cite{Hinkle_2021}. The X-ray light curve showed early time variations on scales corresponds to $\sim 0.01\%$ Eddington luminosity, followed by a flare with a peak of $\sim 10^{43} \rm \, erg\,s^{-1}$, corresponding to $\sim 1 \%$ Eddington luminosity. Due to a seasonal observing gap, the rise time of the X-ray emission is unknown, however, it is evident that there is a significant time delay of $\sim 110$\,days between the X-ray and radio emission peaks.

In parallel to the late-time X-ray flare, the X-ray hardness ratio of AT\,2019azh also varies, from an intermediate state to a completely soft state. Later on, as the late-time X-ray flux declines, the hardness ratio resumes its original intermediate state. Overall, the combination of an X-ray flare followed by a hardness ratio variation resembles the cycle of accretion state transitions observed in XRBs \citep{Fender_2004}. On the other hand, while we observed a late-time radio flare following the X-ray rebrightening, similar to what occurs in XRBs, this radio flare occurs when the X-ray emission is completely soft, which is atypical to radio flares observed in XRBs. 

Changes in the accretion state were also suggested to occur in AGN in a similar manner to the observed changes in XRBs, and the overall X-ray behaviour observed in the TDE AT\,2019azh. Different accretion states in AGN can effect the growth of their central SMBH. Due to their nature, TDEs can also contribute to the growth of this SMBH, however, the role played by TDEs in its evolution is still debated. Different simulations taking into account TDEs \citep{freitag_2002, Pfister_2021} suggest that they could play a key role in the growth of light massive black holes (MBH; $\rm M < 10^{5-6} \, M_{\odot}$) but fewer TDEs are expected when the MBH becomes sufficiently massive to reach the luminosity of an AGN. At that point, the contribution of TDEs to the growth of SMBHs is thought to be negligible. However, if episodes of different accretion rates, as hinted for AT\,2019azh, are more common, TDEs may play a larger role in the growth of AGN. Future work on late-time radio flare and transition in the accretion state of TDEs may shed more light on the role TDEs play in the evolution of SMBH.

The combined properties of the late-time radio and X-ray flares observed in AT\,2019azh are unique. While delayed late-time radio flares have been reported recently in the TDEs ASASSN-15oi \citep{Horesh_2021a}, iPTF\,16fnl \citep{Horesh_2021b}, and IGRJ 12580+0134 \citep{perlman_2021}, a hardness ratio transition of the X-ray emission, associated with the radio flare, has not been observed so far. This does not necessarily mean that such a transition did not occur in these cases, but rather points to the lack of late-time X-ray data in parallel to the radio observations (or vice versa). The case of AT\,2019azh further supports the possibility that (delayed) late-time radio flares are a common phenomena and motivates future late-time simultaneous radio and X-ray observations. This will not only contribute to uncover additional late-time radio flares, but also possibly help explore accretion state transitions around SMBH in real time.

\section*{Acknowledgements}

We thank the anonymous referee for improving this manuscript and T. Piran and N. Stone for useful discussions. A.H. is grateful for the support by the I-Core Program of the Planning and Budgeting Committee and the Israel Science Foundation, and support by ISF grant 647/18. This research was supported by Grant No. 2018154 from the United States-Israel Binational Science Foundation (BSF). We acknowledge the staff who operate and run the AMI-LA telescope at Lord's Bridge, Cambridge, for the AMI-LA radio data. AMI is supported by the Universities of Cambridge and Oxford, and by the European Research Council under grant ERC-2012-StG-307215 LODESTONE. D.R.A.W and J.B were supported by the Oxford Centre for Astrophysical Surveys, which is funded through generous support from the Hintze Family Charitable Foundation. S. Schulze acknowledges support from the G.R.E.A.T research environment, funded by {\em Vetenskapsr\aa det},  the Swedish Research Council, project number 2016-06012.

\software{
CASA \citep{mcmullin_2007},
AMI-LA data reduction package \citep{perrott_2013},
\texttt{emcee} python package \citep{foreman_2013}}.
\bibliography{main.bib}

\end{document}